# Core-Shell Nanofiber Containing Large Amount of Flame Retardants via Coaxial Dual-Nozzle Electrospinning as Battery Separators

Yun Zhao [1,*], Yuqing Chen [1], Yuqiong Kang [1], Li Wang [1], Shaobin Yang [2], Zheng Liang [3,*] and Yanxi Li [3,*]

[1] Institute of Nuclear & New Energy Technology, Tsinghua University, Beijing 100084, China; yzhaozjut@tsinghua.edu.cn
[2] College of Materials Science and Engineering, Liaoning Technical University, Fuxin 123000, China
[3] Department of Materials Science and Engineering, Stanford University, Stanford, CA 94305, USA; lianzhen@stanford.edu
* Correspondence: yzhao.zjut@hotmail.com (Y.Z.); lianzhen@stanford.edu (Z.L.); liyanxi@stanford.edu (Y.L.)

**Abstract:** Lithium-ion batteries have attracted enormous interests recently as promising power sources. However, the safety issue associated with the employment of highly flammable liquid electrolyte impedes the further development of next-generation lithium-ion batteries. Recently, researchers reported the use of electrospun core-shell fiber as the battery separator consisting of polymer layer as protective shell and flame retardants loaded inside as core. In case of a typical battery shorting, the protective polymer shell melts during thermal-runaway and the flame retardants inside would be released to suppress the combustion of the electrolyte. Due to the use of a single precursor solution for electrospinning containing both polymer and flame retardants, the weight ratio of flame retardants is limited and dependent. Herein, we developed a dual-nozzle, coaxial electrospinning approach to fabricate the core-shell nanofiber with a greatly enhanced flame retardants weight percentage in the final fibers. The weight ratio of flame retardants of triphenyl phosphate in the final composite reaches over 60 wt.%. The LiFePO$_4$-based cell using this composite nanofiber as battery separator exhibits excellent flame-retardant property without compromising the cycling stability or rate performances. In addition, this functional nanofiber can also be coated onto commercial separators instead of being used directly as separators.

## 1. Introduction

Lithium-ion battery (LIBs) has been considered as one of the most promising energy-storage devices recently [1]. Although its energy density keeps increasing, battery safety is the prerequisite for the further development and commercial realization [2-8]. The conventional LIBs use flammable organic liquid electrolytes such as ethylene carbonate (EC), diethyl carbonate (DEC), and dimethyl carbonate (DMC), leading to severe safety hazards (for example, fires and explosions). In the case of short circuits, undesirable exothermic reactions from electrolyte decomposition will cause a rapid increase in battery temperature as well as thermal runaway. The high temperature will cause the ignition of the flammable electrolyte, eventually leading to fires or explosions. And this problem will be even magnified with the use of next-generation high energy density LIBs [9-11].

Considerable efforts have been devoted to tackle the aforementioned issues, including the utilization of: (1) non-flammable organic electrolytes or co-solvent [12-18]; (2) high concentration electrolytes [19-21]; (3) electrolytes additives [22-25]. Generally, the non-flammable electrolytes and co-solvents are based on fluorinated, phosphate and ionized liquids, which have high viscosities, low solubility of salts, and are incompatible with common anode materials [19-20], resulting in the poor electrochemical performances when used as LIB electrolytes. Moreover, high concentration electrolytes always suffer from much higher viscosity, lower ionic conductivity and high costs, hindering its practical applications [19]. So far, the most efficient and economical strategy is to add flame-retardant electrolyte additives to lower the risk of fires or explosions [26-28]. However, this improvement of battery safety is on the cost of battery performances since the presence of large amount of flame retardants in the electrolyte would significantly lower the ionic conductivity [29]. Therefore, developing a method to eliminate the trade-off between the safety and the electrochemical performance of the battery is of great importance for further development and practical applications of next-generation LIBs.

Encapsulating flame retardants into separator provides a novel strategy. Recently, Cui et al. has developed a novel electrospun core-shell microfiber structure with protective polymer layer as the shell and triphenyl phosphate (TPP), a popular organophosphorus-based flame retardant confined inside as the core [30]. The polymer shell which is the poly(vinylidene fluoride–hexafluoropropylene) (PVDF-HFP), blocks the direct contact of TPP to electrolyte thus preventing any negative effects of TPP on the electrochemical properties of the electrolyte. While the TPP confined inside would be released to the electrolyte in case of a battery thermal-runaway since PVDF-HFP shell would melt as the temperature increases. The released TPP in the electrolyte further suppress the combustion of flammable organic electrolyte. It is worth noting that the amount of TPP plays a key role on the flame-retardant effects [30]. As shown by Cui et al., the flammability of the electrolyte is dramatically reduced with the increasing concentration of TPP. In addition, for a commercial pouch cell, separator only accounts for around 5 wt.% of the total weight, therefore higher TPP percentage in this composite needs to be achieved [30]. However, due to use of a single-nozzle electrospinning from a single precursor solution in Cui's work, the concentration of TPP is limited [30]. Cui et al. demonstrated the fibrous composite with up to 40 wt.% of TPP [30].

Herein, by using a dual-nozzle, coaxial electrospinning method, we developed a flexible fabrication method of core-shell nanofiber with large quantity of TPP up to more than 60 wt.% encapsulated inside as separators for safe LIBs. In addition, the as-prepared fibrous material displays a very uniform flame retardants distribution, with almost no noticeable agglomeration observed. We further tested and confirmed the flame-retardant capability of the electrospun fiber compared with commercial separator. Finally, the constructed half-cell using lithium iron phosphate (LiFePO$_4$) as cathodes, lithium metal as anode, and this electrospun fiber as separator delivered good cycling and rate performance, with a stable specific capacity of 130 mAh g$^{-1}$ at 1 C. And this indicates the improvement of battery safety is achieved without compromising the battery performances too much.

**2. Materials and Methods**

*2.1. Materials:* Poly(vinylidene fluoride-hexafluoropropylene) (PVDF-HFP, average M$_w$ ≈ 455,000, average M$_n$ ≈ 110,000, pellets), dimethylacetamide (DMAc, 99.8%), N-methyl-2-pyrrolidone (NMP, 99.5%) and acetone (99.9%) were purchased from Sigma-Aldrich. TPP was purchased from Energy Chemical. All of these reagents were used without further purification. Electrolyte (1M lithium hexafluorophosphate (LiPF$_6$) dissolved in a mixture of ethylene carbonate (EC) and dimethyl carbonate (DMC) (v/v = 1:1), moisture ≤ 10 ppm), poly(vinylidene fluoride) (PVDF, 99.5%), lithium metal foil (99.9%), copper foil (12 μm, 99.8%), aluminum foil (16 ± 2 μm, 99.54%), carbon black C45 and coin-type cell CR2032 was purchased from MTI Shenzhen Kejing Star Technology. Two types of commercial separators were used in this work. The commercial separator type I is a Celgard 2320 separator with polypropylene (PP)/polyethylene (PE)/PP trilayer and 20-μm thickness. The commercial separator type II is a Celgard 2500 separator with PP layer and 25-μm thickness.

*2.2. Methods:* Dual nozzle coaxial electrospinning facility was purchased from Changsha Nanoapparatus China, with inner radius 0.18 mm and outer radius 0.59 mm. In order to fabricate desired fibrous materials with high content of flame retardant, the electrospinning precursor solution for polymer shell was prepared by adding 18 wt.% PVDF-HFP in the mixture of DMAc/acetone (w/w = 3:7), then the solution was stirred at room temperature for 6 h, followed by ultrasonic treatment for 30 minutes. Meanwhile, the precursor solution for the core was prepared by dissolving TPP (the mass percentage can be up to more than 60 wt.%) in the mixture of DMAc/acetone (w/w = 3:7). During dual-nozzle coaxial electrospinning, the flow rates of the core and shell solutions were 1.08 mL h$^{-1}$ and 0.54 mL h$^{-1}$, respectively, and the needle-to-collector distance was about 15 cm. A high voltage of 15 kV was applied to the nozzle to start the spinning process and the electrospun fibers were collected on a copper foil coated rotating drum (rotational speed was about 150 r/min). The obtained membrane was washed by ethanol and dried at room temperature.

*2.3. Characterization:* The thermal gravimetric analysis (TGA, Netzsch, STA 409 PC) measurement was carried out in air flow at a heating rate of 10 °C min$^{-1}$. The morphology and elemental analysis of the fibrous material were characterized by transmission electron microscopy (TEM, Hitachi, HT7700), scanning electron microscopy (SEM, Hitachi, SU-8010) and its energy dispersive spectrometer (EDS, Hitachi, SU-8010). X-ray photoelectron spectroscopy (XPS, ULVAC-PHI, PHI Quantro SXM) was used to characterize the as-prepared samples.

*2.4. Electrochemical Characterization:* The electrochemical performance was examined using CR2032 coin-type half-cells assembled in an argon-filled glove box with LiFePO$_4$ as the working electrode and lithium foil as the counter electrode. To prepare the LiFePO$_4$ electrode, LiFePO$_4$, Carbon black C45 and PVDF with a mass ratio of 8:1:1 were dissolved in NMP to form a homogeneous slurry. The slurry was coated onto current collector via a common doctor-blade coating method, which was dried in a vacuum oven at 120 °C for 2h. The electrolyte is 1M LiPF$_6$ in a mixture of ethylene carbonate (EC) and dimethyl carbonate (DMC) (v/v = 1:1). The galvanostatic discharge-charge cycling was performed in land system (CT2001A) over a range of applied voltage of 2.5-4.2 V at a constant C-rate of 0.25 C in the first 3 cycles for activation and at 1 C in the following cycles.

*2.5. Ignition Experiments:* For the ignition experiments, firstly, all the separators/fiber networks were folded four times and clamped with glass plates. Then the electrospun fibers were heated in air from room temperature to 180 °C to release the flame retardants inside. After cooling to room temperature, these fiber networks were wetted by a flammable electrolyte (1 M LiPF$_6$ in a mixture of EC and DMC (v/v = 1:1)).

**3. Results**

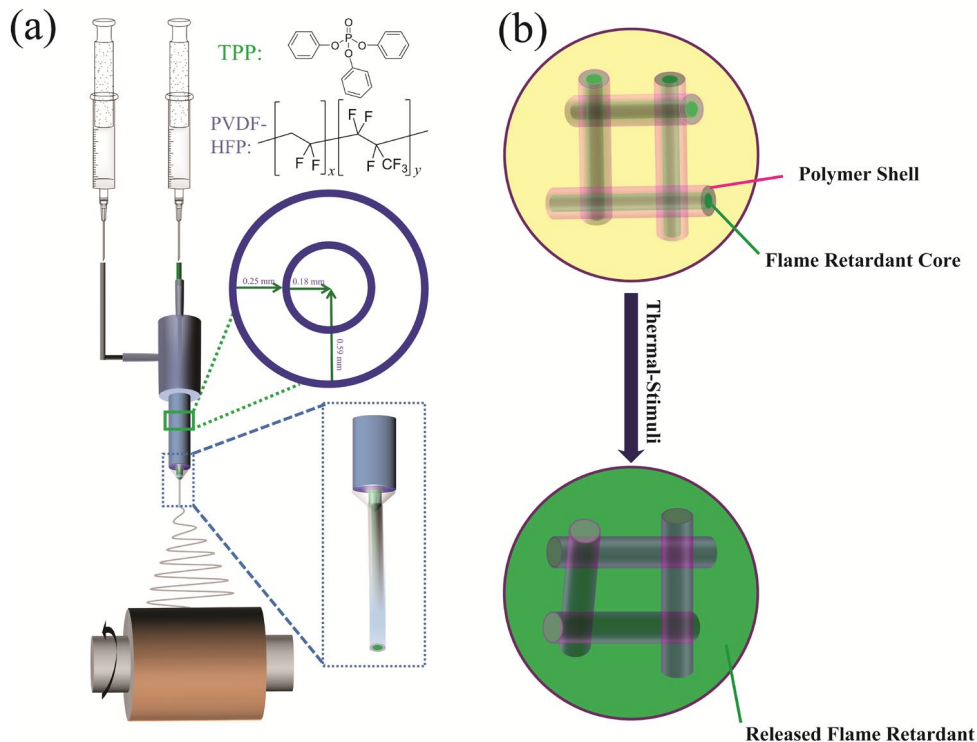

**Figure 1.** Schematic illustration of the coaxial electrospinning process using a dual nozzle. (**a**) The TPP and PVDF-HFP solutions were injected by syringes into the core and shell channels of the nozzle, respectively. The Taylor cone region was magnified as shown in the box. (**b**) The as-prepared electrospun fiber consists of TPP as the core and PVDF-HFP as the shell. The confined flame retardants inside the core will be released to dissolve in the electrolyte and suppress the combustion.

The fabrication process for our proposed core-shell nanofiber with high content of flame retardants was vividly presented in **Figure 1**. The dual-nozzle coaxial electrospinning technique was chosen to prepare the core-shell structure. PVDF-HFP as the protective layer was dissolved in a mixture of dimethylacetamide (DMAc)/acetone (w/w = 3:7) while another mixture solution containing TPP was injected into the core channel of the dual nozzle. During electrospinning, TPP solution, as the central part, can be extruded through the inner capillary while the spinnable PVDF-HFP polymer solution is extruded through the outer capillary. The protective polymer is firstly precipitated from the outer solution such that the TPP can be uniformly dispersed and encapsulated in the inner region (**Figure 1a**). It was worth noting that allowing the shell solution to be extruded firstly was very important to establish a stable Taylor cone for the desired core-shell structure. With a stable Taylor cone, the inner TPP solution could not leak easily to form agglomeration or lead to non-homogeneous distribution in fibers (**Figure S1**). After the solutions sprayed out from the nozzle tip, as the core and shell solution in contact, PVDF-HFP would be precipitated immediately as a protection layer at the interface, thus preventing the TPP to diffuse from inner to outer solution.

Morphology of the as-prepared electrospun fibers with high content of flame retardants was examined by scanning electron microscope (SEM) as shown in **Figure 2a-b**. Though the diameter of the electrospun fibers ranges from 200 nm to 500 nm, these fibers are very smooth and show almost no agglomeration (**Figure 2a-b**). In addition, the as-prepared fiber network exhibits excellent flexibility, which was demonstrated by rolling up on a pen and folding several times without any cracks or structural damages (**Figure S2**).

In order to confirm the core-shell structure of our electrospun fibers (noted as TPP@PVDF-HFP), transmission electron microscope (TEM), energy-dispersive X-ray spectroscopy (EDX) and X-ray photoelectron spectroscopy (XPS) were utilized. As illustrated in **Figure 2d**, the boundary of the core region and shell region can be clearly observed, indicating the successful encapsulation of TPP core in PVDF-HFP shell. According to our calculation (**Figure S3**), the polymer shell thickness differs not too much with our observations from the TEM image (**Figure 2d**). To examine the distribution of flame retardants inside the fiber, EDX and XPS were utilized. The EDX elemental mapping shows the homogeneous distribution of C, F and P, implying uniform distribution of TPP inside PVDF-HFP (**Figure 2e**). Furthermore, the coexistence of C and F elements from the XPS spectra (**Figure 2f**) in the composite fiber indicated the existence of PVDF-HFP as the shell, while no peaks of P element was detected, suggesting the successful encapsulation of TPP inside as the core (**Figure 2f**). In addition, in case of a battery thermal runaway, the outer layer of the fiber network (PVDF-HFP) melts and the inner TPP released into the electrolyte (**Figure 2c**), and this can be confirmed from the appearance of characteristic peaks of elemental P in the XPS spectrum after thermal runaway (**Figure S4**), which further proves the core-shell structure of the as-prepared TPP@PVDF-HFP composite fiber.

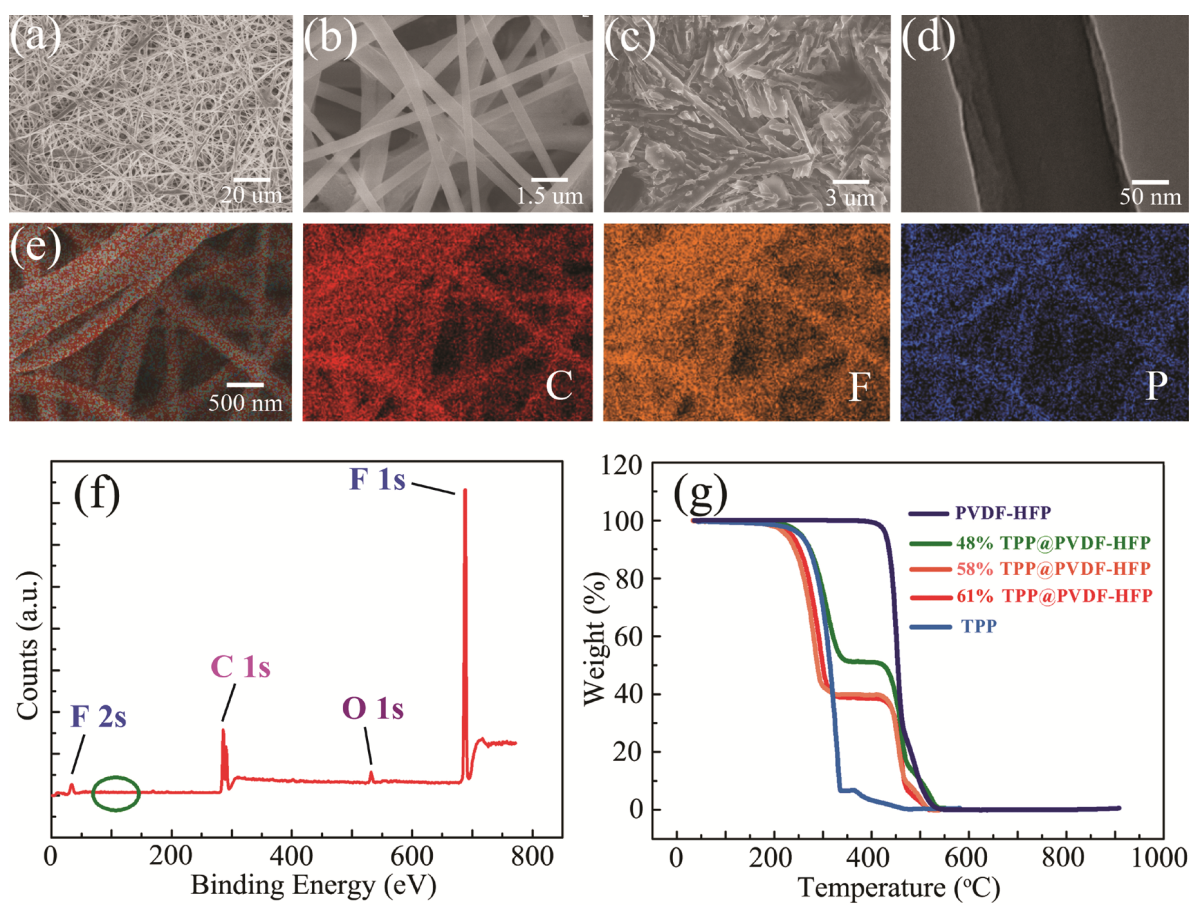

**Figure 2.** The morphology and compositional characterization of the coaxial electrospun nanofibers. (**a**) Magnified and (**b**) low-magnification SEM images of the TPP@PVDF-HFP nanofibers. (**c**) The morphology of TPP@PVDF-HFP fibers after thermal-runaway. (**d**) TEM image of the TPP@PVDF-HFP electrospun core-shell fiber. (**e**) The EDS mapping results of the TPP@PVDF-HFP fibers. (**f**) The XPS data of the TPP@PVDF-HFP nanofibers before thermal-runaway. (**g**) TGA results of TPP@PVDF-HFP nanofibers with different amount of flame retardants confined inside.

The contents of flame retardant in the electrospun fibers were investigated by thermogravimetric analysis (TGA). Herein, we prepared electrospun core-shell fibers with different amount of flame retardants contained inside by adjusting the concentration of core TPP solution. Theoretically, the concentration of TPP in the core solution can be related to the overall concentration of TPP in the entire composite fiber according to our calculations. Specifically, the TPP concentration in the core solution of 33.3 wt.%, 50.0 wt.% and 66.6 wt.% can be correlated to the final composite fibers with 48 wt.%, 58 wt.% and 65 wt.% weight percentage of TPP (**Figure S5**). And this calculation matches perfectly with the experimental observation as showed in **Figure 2g** from the TGA analysis. The TPP showed an obvious weight loss in the range of 250 °C to 330 °C while the decomposition of PVDF-HFP occurred at around 425 °C. For our coaxial electrospun fibers with TPP confined in the PVDF-HFP shell, there are two stages for the weight loss in relevant to TPP and PVDF-HFP decomposition, respectively, demonstrating the successful encapsulation of TPP inside.

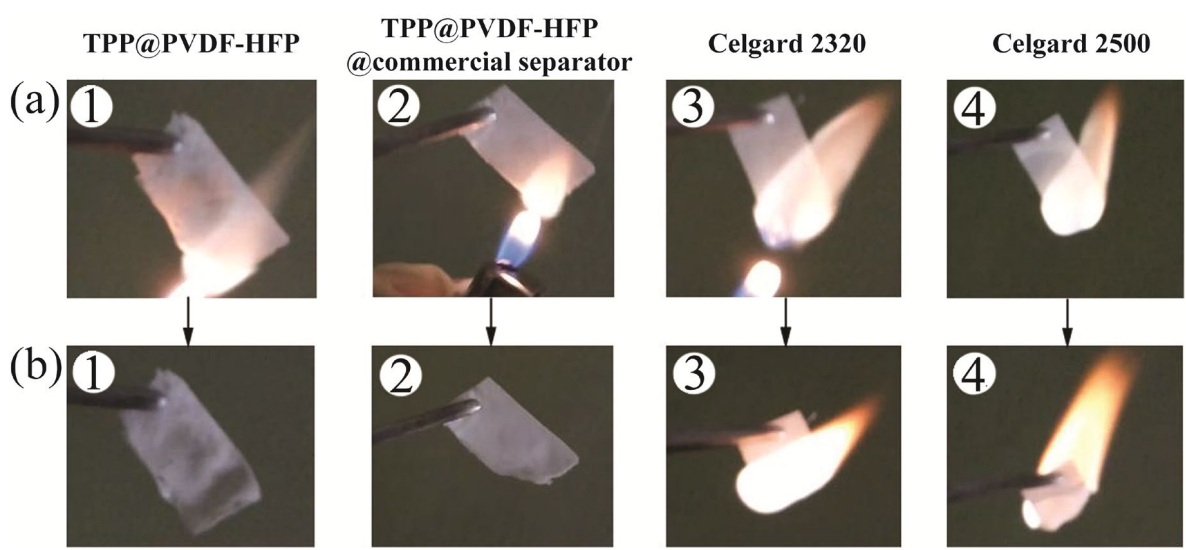

**Figure 3.** Digital images showing the examination of flame-retardant properties of various separators/fiber networks by ignition experiment. Images of (**a1-a4**) display the status just as the ignition starts for TPP@PVDF-HFP fiber network, TPP@PVDF-HFP@commercial separator, commercial Celgard 2320 separator and commercial Celgard 2500 separator, respectively. While images of (**b1-b4**) represent the corresponding status after ignition.

In order to prove the flame-retardant property of the as-prepared core-shell composite fibers, ignition experiments were carried out on our TPP@PVDF-HFP composite fibers compared with two types of commercial separators (Celgard 2320 and Celgard 2500) (**Figure 3a-b**). In addition, commercial separator (Celgard 2320) was also coated (**Figure S6**) by TPP@PVDF-HFP composite fibers (denoted TPP@PVDF-HFP@commercial separator) and subjected to the ignition experiments (**Figure 3a-b**). These separators/fiber networks were wetted by liquid electrolyte of 1 M lithium hexafluorophosphate (LiPF$_6$) in the mixture of EC and DMC (v/v = 1:1), which is the real LIB electrolyte for a pouch cell and then subjected to an ignition. When heated to 180 °C, the PVDF-HFP fibers was broken and TPP would be released. Therefore, when wetted by EC/DMC electrolyte, the flame retardant (TPP) immediately dissolved into electrolyte, converting the flammable electrolyte into a nonflammable liquid. As can be seen from **Figure 3b**, both TPP@PVDF-HFP and

TPP@PVDF-HFP@commercial separator could not be ignited by a lighter, even for several times (**Figure 3b1-b2**). However, a direct flame could be observed for the commercial separators (two types) with the same electrolyte (**Figure 3b3-b4**). The full process of ignition experiments for the above four types of fibrous materials can be found in **movie S1-S4**, respectively. Apparently, coaxial electrospun core-shell fibers with flame retardants contained inside could greatly improve the battery safety and suppress the combustion.

The effects of the TPP@PVDF-HFP (30 μm) and TPP@PVDF-HFP@commercial separator (35 μm) on the electrochemical performances of LIBs when used as battery separators were examined in the following section. **Figure 4a** displays the specific discharge capacity as a function of the cycle number in LiFePO$_4$ cells at 0.5 C. While the cell with TPP@PVDF-HFP as separator exhibited higher specific capacity than that of TPP@PVDF-HFP@commercial separator, both the cells showed good electrochemical cycling stability during the repeating charge and discharge. In addition, both cells could deliver a stable capacity of more than 120 mAh/g on average for over 100 cycles, which is comparable to LiFePO$_4$ cells with normal separators [35]. This good cycling behavior further confirmed that the flame retardants (TPP) was encapsulated inside the PVDF-HFP shell, otherwise the cycling performance would be greatly negatively affected. Cells with these two separators were further subjected to a C-rate test to investigate the rate capability (**Figure 4b**). Both cells exhibited good rate capabilities, indicating the introduction of the encapsulated TPP has little negative effects on the rate capability. Moreover, with the increase of discharge rate, the Coloumbic efficiency of each cycle gradually declined when using TPP@PVDF-HFP as separator (**Figure 4c**).

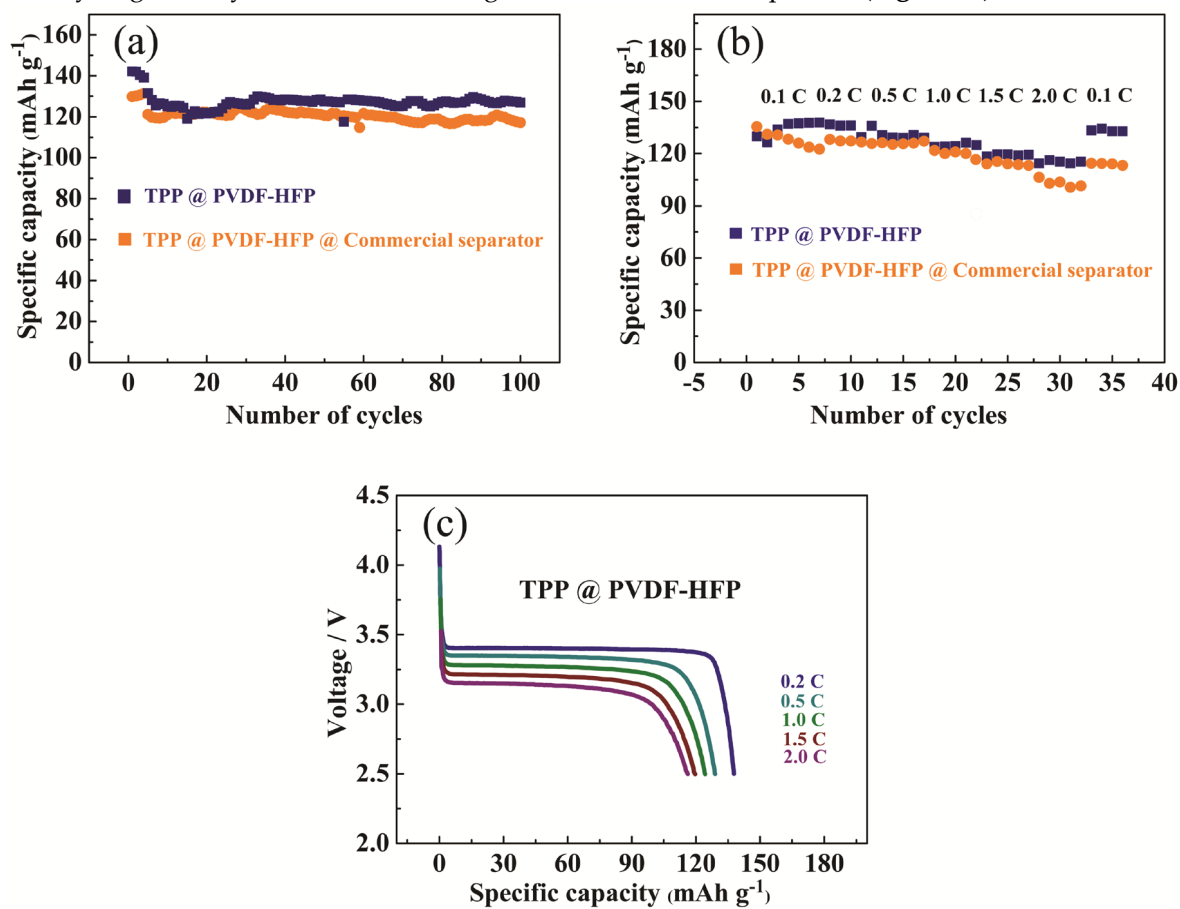

**Figure 4.** Electrochemical performances of the LiFePO$_4$ cells using the as-prepared coaxial electrospun fibers as separators. (**a**) Cycling of the LiFePO$_4$ half cells over 2.5-4.2 V at a constant C-rate of 0.25 C in the first 3 cycles for activation and at 1 C in the following cycles. (**b**) Rate performances under the same conditions. (**c**) The discharge voltage profiles at 0.2 C to 2.0 C.

## 4. Conclusions

In conclusion, we have fabricated an electrospun fibrous material which could either be directly adopted as flame-retardant separator or be coated on a commercial separator for LIBs. This electrospun nanofiber network has a thermal-triggered flame-retardant property for LIBs such that in case of a battery thermal-runaway, it will release its flame-retardant core component out and suppress the combustion of the organic electrolyte. Unlike the previous methods to fabricate the flame retardants containing fibers based on single-nozzle electrospinning from a single precursor solution, our proposed approach involves dual-nozzle coaxial electrospinning out of two different precursor solutions, with one for polymer shell and the other for flame retardants. As a result, the concentration of the flame retardants in the precursor solution is independent of the polymer solution and can be tuned to a relatively high level. The as-prepared core-shell fibers contain a large quantity, more than 60 wt.% of flame retardants out of the entire composite fiber, which is a significant improvement compared with previous versions. Since the nonflammability of the electrolyte would increase drastically with the increasing flame retardants weight ratio, our developed electrospun fibers with more than 60 wt.% enable excellent flame-retarding property. In addition, during normal battery cycling, the outer protective shell confined flame retardants inside and prevent the exposure and dissolution of flame retardants into liquid electrolyte, which otherwise would have negative influence on battery performances. Therefore, when used as battery separators, our developed coaxial electrospun core-shell fibers reduce risks of safety hazards without compromising the battery performance.

**Supplementary Materials:** The following are available online. **Figure S1.** The morphology of the as-prepared TPP@PVDF-HFP when the Taylor cone was unstable. **Figure S2.** The test of the flexibility of TPP@PVDF-HFP fiber network by (**a**) rolling up on a pen or (**b**) after folding several times. **Figure S3.** The calculation of dimensions for a single fiber of TPP@PVDF-HFP fiber network. **Figure S4.** The XPS data of the TPP@PVDF-HFP fiber network after thermal stimuli. **Figure S5.** The theoretical calculation of overall TPP mass fraction in the TPP@PVDF-HFP fiber network. **Figure S6.** The SEM image of electrospun nanofibers on the commercial Celgard 2320 separator denoted as TPP@PVDF-HFP@commercial separator. **Movie. S1** Full process of ignition experiments on TPP@PVDF-HFP fibers. **Movie. S2** Full process of ignition experiments on TPP@PVDF-HFP@commercial separators fibers. **Movie. S3** Full process of ignition experiments on commercial separator type I (Celgard 2320). **Movie. S4** Full process of ignition experiments on commercial separator type II (Celgard 2500).

**Author Contributions:** conceptualization, Y.Z., L.W. and Z.L.; investigation, Y.L.; data curation, Y.C. and Y.K.; writing—original draft preparation, Y.Z. and Z.L.; writing—review and editing, Z.L. and S.Y.; supervision, Y.Z.;

**Funding:** The work was supported by the National Natural Science Foundation of China (No. U1564205), Ministry of Science and Technology of China (No. 2013CB934000, 2016YFE0102200).

**Acknowledgments:** The authors acknowledge the support from ChemDraw software online free version for completing the Figure 1.

**Conflicts of Interest:** The authors declare no conflict of interest.

## References

[1] Miao, Y.; Hynan, P.; Jouanne, A.; Yokochi, A. Current Li-Ion Battery Technologies in Electric Vehicles and Opportunities for Advancements. *Energies* **2019**, *12*, 1074.
[2] Chen, Z.; Ren, Y.; Lee, E.; Johnson, C.; Qin, Y.; Amine, K. *Adv. Energy Mater.* **2013**, *3*, 729-736.
[3] Chen, J. Recent Progress in Advanced Materials for Lithium Ion Batteries. *Materials* **2013**, *6*, 156–183.
[4] Chen, Z.; Hsu, P. C.; Lopez J.; Li, Y. Z.; To, J. W. F.; Liu, N.; Wang, C.; Andrews, S. C.; Liu, J.; Cui, Y.; Bao, Z. N. *Nature Energy* **2016**, *1*, 15009.
[5] Zhang, S.; Cao, J.; Shang, Y. M.; Wang, L.; He, X. M.; Li, J. J.; Zhao, P.; Wang, Y. W. *J. Mater. Chem. A* **2015**, *3*, 17697-17703.
[6] Zhao, P.; Yang, J. P.; Shang, Y. M.; Wang, L.; Fang, M.; Wang, J. L.; He, X. M. *Journal of Energy Chemistry* **2015**, *24*, 138-144.
[7] Yang, J. P.; Zhao, P.; Shang, Y. M.; Wang, L.; He, X. M.; Fang, M.; Wang, J. L. *Electrochimica Acta* **2014**, *121*,

264-269.
[8] H. Li, D. B. Wu, J. Wu, L. Y. Dong, Y. J. Zhu, X. L. Hu, *Adv. Mater.* **2017**, *29*, 1703548.

[9] Haregewoin, A. M.; Wotangoa, A. S.; Hwang, B. J. *Energy Environ. Sci.* **2016**, *9*, 1955-1988.
[10] Adachi M.; Tanaka, K.; Sekai, K. *J. Electrochem. Soc.* **1999**, *146*, 1256-1261.
[11] Fei, H. F.; An, Y. L.; Feng, J. K.; Ci, L. J.; Xiong, S. L. *RSC Adv.* **2016**, *6*, 53560-53565.
[12] K. Xu, *Chem. Rev.* **2004**, *104*, 4303-4418.
[13] K. Xu, *Chem. Rev.* **2014**, *114*, 11503-11618.
[14] Suo, L. M.; Borodin, O.; Gao, T.; Olguin, M.; Ho, J.; Fan, X. L.; Luo, C.; Wang, C. S.; Xu, K. *Science* **2015**, *350*, 938,
[15] Wang, F.; Suo, L. M.; Liang, Y. J.; Yang, C. Y.; Han, F. D.; Gao, T.; Sun W.; Wang, C. S. *Advanced Energy Materials* **2017**, *7*, 1600922.
[16] Suo, L. M.; Borodin, O.; Sun, W.; Fan, X. L.; Yang, C. Y.; Wang, F.; Gao, T.; Ma, Z. H.; Schroeder, M.; Cresce, A. V.; Russell, S. M.; Armand, M.; Angell, A.; Xu, K.; Wang, C. S. *Angew. Chem. Int. Ed.* **2016**,*55*,7136-7141.
[17] Kalhoff, J.; Eshetu, G. G.; Bresser, D.; Passerini, S. *ChemSusChem* **2015**, *8*, 2154-2175.
[18] Armand, M.; Endres, F.; MacFarlane, D. R.; Ohno, H.; Scrosati, B. *Nature Materials* **2009**, *8*, 621-629.
[19] Zhang, H.; Zhou, M. Y.; Lin, C. E.; Zhu, B. K. *RSC Adv.* **2015**, *5*, 89848-89860.
[20] Dai, J. H.; Shi, C.; Li, C.; Shen, X.; Peng, L. Q.; Wu, D. Z.; Sun, D. H.; Zhang P.; Zhao, J. B. *Energy Environ. Sci.* **2016**, *9*, 3252-3261.
[21] Sh, C.; Zhang, P.; Huang, S. H.; He, X. Y.; Yang, P. T.; Wu, D. Z.; Sun, D. H.; Zhao, J. B. *J. Power Sources* **2015**, *298*, 158-165.
[22] Liu, W.; Lee, S. W.; Lin, D. C.; Shi, F. F.; Wang, S.; Sendek, A. D.; Cui, Y. *Nature Energy* **2017**, *2*, 17035.
[23] Lin, D. C.; Liu, W.; Liu, Y. Y.; Lee, H. R.; Hsu, P. C.; Liu, K.; Cui, Y. *Nano Lett.* **2016**, *16*, 459-465.
[24] Yue, L. P.; Ma, J.; Zhang, J. J.; Zhao, J. W.; Dong, S. M.; Liu, Z. H.; Cui, G. L.; Chen, L. Q. *Energy Storage Materials* **2016**, *5*, 139-164.
[25] Wang, L.; Li, N.; He, X. M.; Wan, C. R.; Jiang, C. Y. *J. Electrochem. Soc.* **2012**, *159*, A915-A919.
[26] Rectenwald, M. F.; Gaffen, J. R.; Rheingold, A. L.;Morgan, A. B.; Protasiewicz, J. D. *Angew. Chem. Int. Ed.* **2014**, 53, 4173-4176.
[27] Xiang, H. F.; Xu, H.Y.; Wang, Z. Z.; Chen, C. H. *J. Power Sources* **2007**, *173*, 562-564.
[28] Zhou, D. Y.; Li, W. S.; Tan, C. L.; Zuo, X. X.; Huang, Y. J. *J. Power Sources* **2008**, *184*, 589-592.
[29] Zeng, Z. Q.; Jiang, X. Y.; Wu, B. B.; Xiao, L. F.; Ai, X. P.; Yang, H. X.; Cao, Y. L. *Electrochimica Acta* **2014**, *129*, 300-304.
[30] Liu, K.; Liu, W.; Qiu, Y. C.; Kong, B.; Sun, Y. M.; Chen, Z.; Zhuo, D.; Lin, D. C.; Cui, Y. *Sci. Adv.* **2017**, *3*, e1601978.
[31] Cho, T. H.; Tanaka, M.; Onishi, H.; Kondo, Y.; Nakamura, T.; Yamazaki, H.; Tanase, S.; Sakai, T. *J. Power Sources* **2008**, *181*, 155-160.
[32] Xiao, K.; Zhai, Y. Y.; Yu, J. Y.; Ding, B. *RSC Adv.* **2015**, *5*, 55478-55485.
[33] Teo, W. E.; Ramakrishna, S. *Nanotechnology*, **2006**, *17*, R89.
[34] Wang, M.; Fang, D. W.; Wang, N. N.; Jiang, S.; Nie, J.; Yu, Q.; Ma, G. P. *Polymer* **2014**, *55*, 2188-2196.
[35] Satyavani, T.V.S.L.; Kumar, A.S.; Rao, P.S.V.S. *Engineering Science and Technology, an International Journal* **2016**, *19*, 178-188.

# Supplementary Materials

# Core-Shell Nanofiber Containing Large Amount of Flame Retardants via Coaxial Dual-Nozzle Electrospinning as Battery Separators

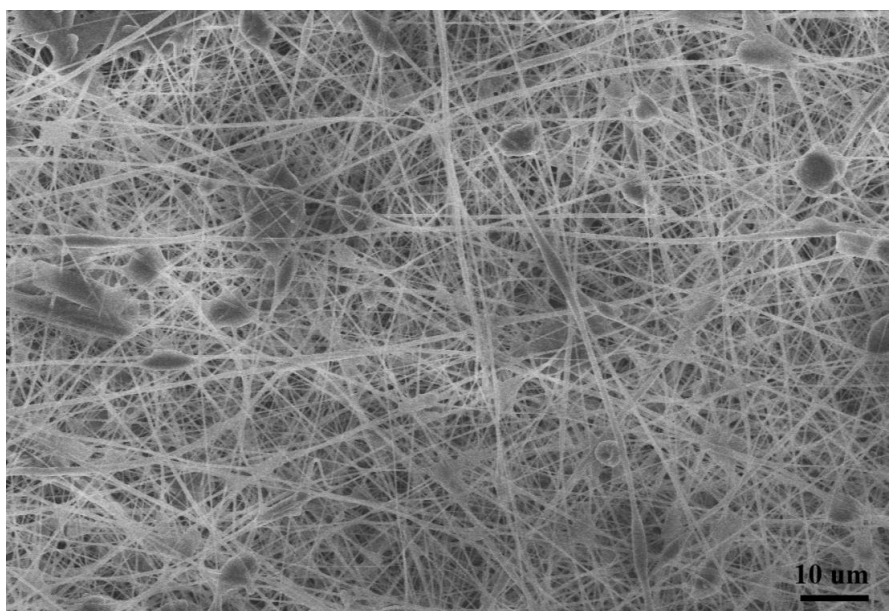

**Figure S1.** The morphology of the as-prepared TPP@PVDF-HFP when the Taylor cone was unstable.

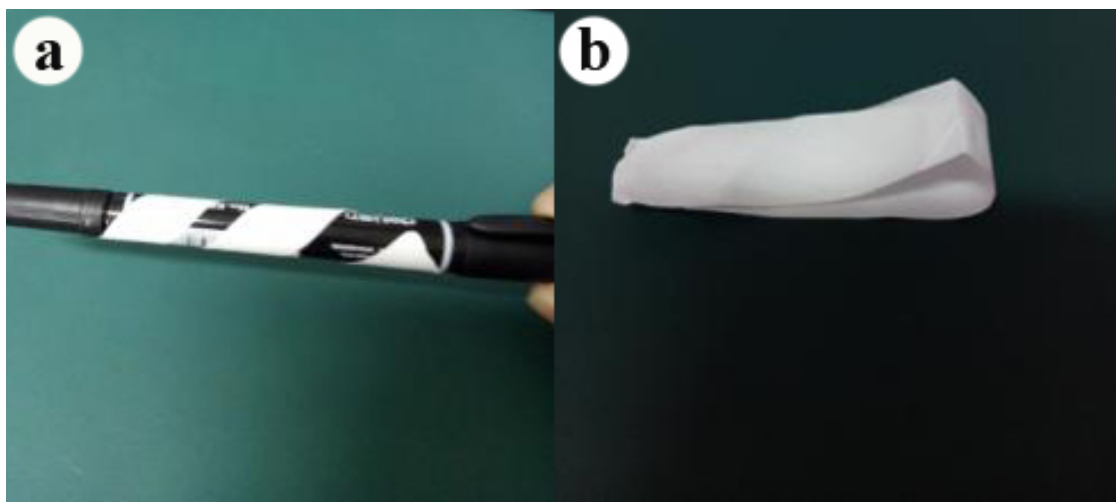

**Figure S2.** The test of the flexibility of TPP@PVDF-HFP fiber network by (a) rolling up on a pen or (b) after folding several times.

**Inner nozzle:** TPP solution

flow rate $V_1 = 0.54$ mL h$^{-1}$

concentration $\omega_1 = 66.7$ wt.%

solution density $\rho_1$

TPP density $\rho_{TPP} = 1.21$ g cm$^{-3}$

TPP mass $m_{TPP}$

**Outer nozzle:** PVDF-HFP solution

flow rate $V_2 = 1.08$ mL h$^{-1}$

concentration $\omega_2 = 18$ wt.%

solution density $\rho_2 \approx \rho_1$

PVDF-HFP density $\rho_{PVDF-HFP} = 1.77$ g cm$^{-3}$

PVDF-HFP mass $m_{PVDF-HFP}$

when forming fibers with core-shell structure, the core radius is $R_1$ and the shell radius is $R_2$.

at the same time, for example 1h:

$m_{TPP} = V_1 \times 1h \times \omega_1 \times \rho_1 = \pi R_1^2 \times 1h \times \rho_{TPP} \times \rho_1$

$m_{PVDF-HFP} = V_2 \times 1h \times \omega_2 \times \rho_2 = \pi R_2^2 \times 1h \times \rho_{PVDF-HFP} \times \rho_2$

Hence, $R_2 = 1.17 R_1$

if $R_2 = 180$ nm, $R_1 = 153$ nm

**Figure S3.** The calculation of dimensions for a single fiber of TPP@PVDF-HFP fiber network.

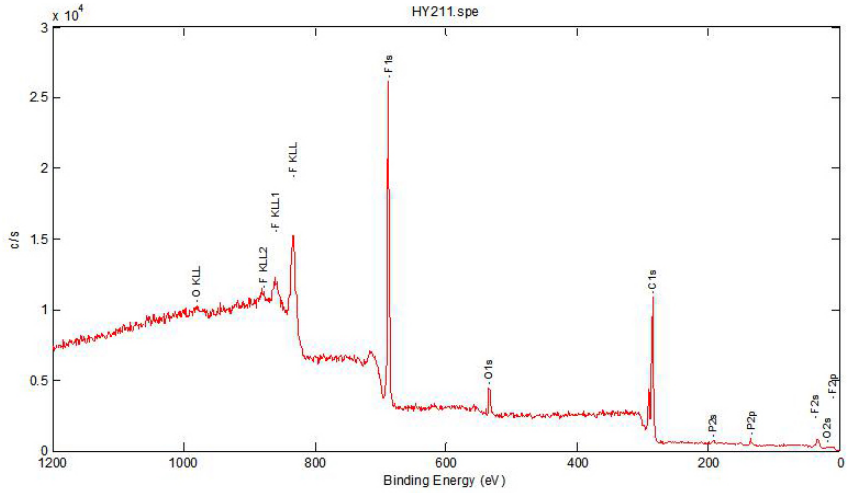

**Figure S4.** The XPS data of the TPP@PVDF-HFP fiber network after thermal stimuli.

**Inner nozzle:** TPP solution

flow rate $V_1 = 0.54$ mL h$^{-1}$

concentration $\omega_1 = 66.7$ wt.%, 50 wt.%, 33.3 wt.%

solution density $\rho_1$

**Outer nozzle:** PVDF-HFP solution

flow rate $V_2 = 1.08$ mL h$^{-1}$

concentration $\omega_2 = 18$ wt.%

solution density $\rho_2 \approx \rho_1$

at the same time, for example 1h:

$m_{TPP} = V_1 \times 1h \times \omega_1 \times \rho_1$

$m_{PVDF-HFP} = V_2 \times 1h \times \omega_2 \times \rho_2$

$\omega_{TPP} = m_{TPP}/(m_{TPP} + m_{PVDF-HFP})$

if $\omega_1 = 66.7$ wt.%

the $\omega_{TPP} = 65.0$ wt.%

**Figure S5.** The theoretical calculation of overall TPP mass fraction in the TPP@PVDF-HFP fiber network.

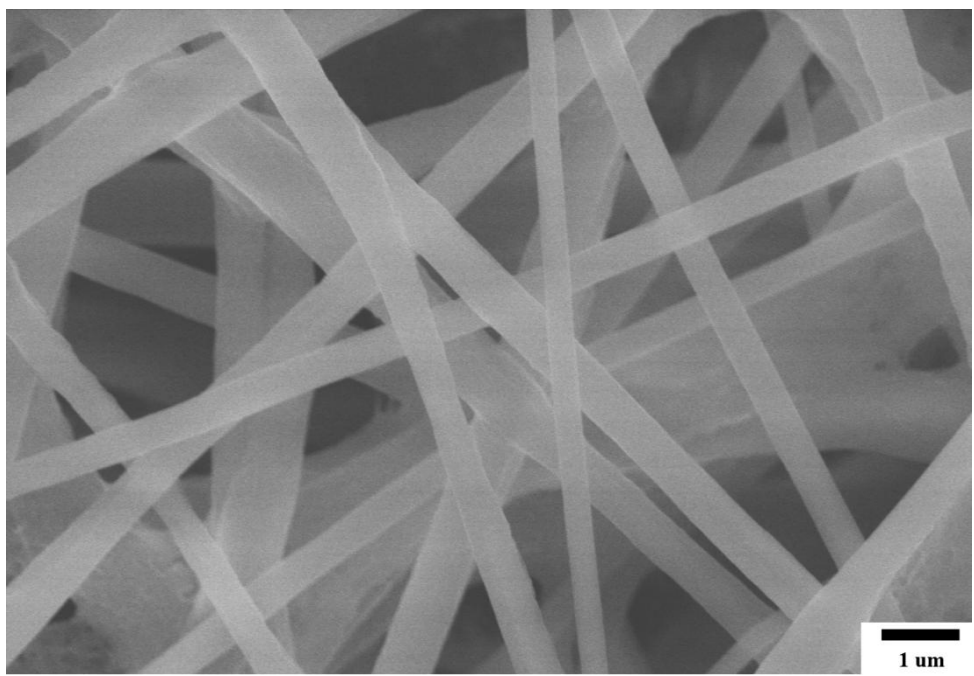

**Figure S6.** The SEM image of electrospun nanofibers on the commercial Celgard 2320 separator denoted as TPP@PVDF-HFP@commercial separator.

**Movies showing the ignition process of various fibrous materials can be found at the link:**
**https://www.dropbox.com/sh/r4zkt962c6ftasq/AABjOpTvhXmiCptYW6feMeAja?dl=0**

**Movie. S1**

Full process of ignition experiments on TPP@PVDF-HFP fibers.

**Movie. S2**

Full process of ignition experiments on TPP@PVDF-HFP@commercial separators fibers.

**Movie. S3**

Full process of ignition experiments on commercial separator type I (Celgard 2320).

**Movie. S4**

Full process of ignition experiments on commercial separator type II (Celgard 2500).